\documentclass[aps,prl,twocolumn,showpacs]{revtex4}

\newcommand{\nc}{\newcommand}
\nc{\gtwid}{\mathrel{\raise.3ex\hbox{$>$\kern-.75em\lower1ex\hbox{$\sim$}}}}
\nc{\ltwid}{\mathrel{\raise.3ex\hbox{$<$\kern-.75em\lower1ex\hbox{$\sim$}}}}
\nc{\comp}{{\rm C}\llap{\vrule height7.1pt width1pt depth-.4pt\phantom t}}

\begin{document}

\preprint{UFIFT-HEP-03-15,CRETE-03-10}

\title{General plane wave mode functions for scalar-driven cosmology}

\author{N. C. Tsamis}
\email[]{tsamis@physics.uoc.gr}
\affiliation{Department of Physics, University of Crete, 
             GR-71003 Heraklion, Greece}

\author{R. P. Woodard}
\email[]{woodard@phys.ufl.edu}
\affiliation{Department of Physics, University of Florida,
             Gainesville, FL 32611, USA}

\date{\today}

\begin{abstract}

We give a solution for plane wave scalar, vector and tensor mode functions
in the presence of any homogeneous, isotropic and spatially flat cosmology
which is driven by a single, minimally coupled scalar. The solution is
obtained by rescaling the various mode functions so that they reduce, with a 
suitable scale factor and a suitable time variable, to those of a massless, 
minimally coupled scalar. We then express the general solution in terms of
co-moving time and the original scale factor.

\end{abstract}


\maketitle

{\it 1. Introduction:} On the largest scales the universe is homogeneous 
and isotropic \cite{KT}. When results from the WMAP satellite are combined 
with other data the energy density of the universe as a fraction of the 
critical density is found to be $\Omega_{tot} = 1.02 \pm .02$ \cite{WMAP}. 
This is consistent with the high degree of spatial flatness predicted by 
most models of inflation \cite{Linde}. It is therefore safe to assume that 
the invariant element relevant to cosmology takes the form,
\begin{equation}
ds^2 = -dt^2 + a^2\!(t) d\vec{x} \cdot d\vec{x} \; . \label{HandI}
\end{equation}

The scale factor $a(t)$ is not observable but the ratio of its current value 
to its value at past time $t$ gives the cosmological redshift (plus one)
experienced by light emitted at that time and received now,
\begin{equation}
z \equiv \frac{a_0}{a(t)} - 1 \; .
\end{equation}
It is common to use redshift as a time variable in cosmology, even for 
epochs from which we detect no radiation. For example, the cosmic 
microwave radiation was emitted within about 100 redshifts before and 
after $z_{dec} = 1089$, and the transition from radiation domination to matter
domination occurred at $z_{eq} \sim 3200$ \cite{WMAP}. Primordial inflation
is conjectured to have ended at $z \gtwid e^{50}$.

The scale factor's logarithmic derivative defines the Hubble parameter,
\begin{equation}
H(t) \equiv \frac{\dot{a}}{a} \; .
\end{equation}
Its physical meaning is the rate at which distant matter is receding due to
the expansion of the universe. Combining WMAP with other data sets determines
its current value to be, $H_0 = (71 {+4 \atop -3})~\frac{\rm km}{\rm s\ Mpc}$ 
\cite{WMAP}. The weak energy condition implies that the Hubble parameter can 
never increase. One typically assumes it has fallen to $H_0$ from an initial 
period of inflation during which it was up to 55 orders of magnitude larger. 

The scale factor's second time derivative defines the deceleration parameter,
\begin{equation}
q(t) \equiv - \frac{a\ddot{a}}{ \dot{a}^2} = -1 - \frac{\dot{H}}{H^2} \; .
\end{equation}
It is less well measured than $H(t)$, but Hubble plots of Type Ia supernovae 
at high redshift are consistent with a current value of $q_0 \sim -.6$ 
\cite{SNIa}. This phase of negative deceleration is a relatively late 
phenomenon, having set in about $z \sim 1$. Before this it was near the 
value $q = +\frac12$ characteristic of a matter dominated universe. At
redshifts much larger than $z_{eq} \sim 3200$ the deceleration parameter was 
near the value $q = +1$ of a perfectly radiation dominated universe. And 
during the conjectured epoch of primordial inflation ($z \gtwid e^{50}$) it 
would have been near the value $q = -1$.

The fact that the deceleration parameter has changed so much over the course
of cosmic history is frustrating because many of the phenomena which 
we can observe from the epoch of matter domination ($q = +\frac12$) are 
believed to have had their origins in quantum fluctuations during the 
epoch of primordial inflation ($q = -1$) \cite{Slava}. The various effects
can be understood using free quantum field theory but the mode functions are 
known only for the case of constant $q(t)$. Hence one must employ either 
perturbative or numerical methods to connect the inflationary mode functions 
--- whose normalization is fixed --- with their matter dominated descendants 
whose effects can be observed \cite{MFB,LL,Rocky}.

The purpose of this paper is to present an exact solution for the scalar, 
vector and tensor mode functions of gravity plus a single, minimally coupled 
scalar with any sort of potential. That we can do this derives from previous 
work \cite{TW1} in which an exact solution was obtained, for any $a(t)$, for 
the mode functions of a massless, minimally coupled scalar,
\begin{equation}
\frac{\partial^2 u(t,k)}{\partial t^2} + H^2\!(t) \Bigl[x^2\!(t,k) + \frac32 
q(t) - \frac34 \Bigr] u(t,k) = 0 \; . \label{MMCSeqn}
\end{equation}
The dimensionless parameter $x(t,k) \equiv k/aH$ is the physical wave 
number expressed in Hubble units. We seek a solution ({\ref{MMCSeqn}) which 
is conventionally normalized,
\begin{equation}
u(t,k) \dot{u}^*\!(t,k) - \dot{u}(t,k) u^*\!(t,k) = i \; , \label{unorm}
\end{equation}
and initially (at $t = t_i$) positive frequency. The desired solution takes 
the form of a row vector containing the two Bessel functions of order $\nu(t) 
\equiv \frac12 \!- q^{\!-\!1}\!(t)$, which comprise (most of) the solution 
for constant $q(t)$, multiplied into a $2 \times 2$ matrix that mixes the two 
solutions whenever $q(t)$ evolves,
\begin{eqnarray}
\lefteqn{u(t,k) \! \equiv \! \sqrt{\frac{\pi a}{2k}} \Bigl(-\frac{x}{2q}
\Bigr)^{\frac12} \Bigl(\!-i {\rm csc}(\nu \pi) J_{\!-\!\nu}({\scriptstyle 
\!-\! \frac{x}{q}}) \, , \, J_{\nu}({\scriptstyle \!-\! \frac{x}{q}}) 
\Bigr) } \nonumber \\
& & \hspace{2.5cm} \times {\cal M}(t,t_i,k) \left(\matrix{1 \cr 1 \!+\! i
{\rm cot}[\nu(t_i) \pi] }\right) . \qquad \label{genu}
\end{eqnarray}
The transfer matrix ${\cal M}(t,t_i,k)$ is the time-ordered product of the 
exponential of a line integral,
\begin{eqnarray}
\lefteqn{{\cal M}(t,t_i,k) \equiv P\left\{ \exp\left[ \int_{t_i}^t dt' 
{\cal A}(t',k) \right] \right\} \; ,} \label{M1} \\
& & \!\!\equiv \sum_{n=0}^{\infty} \int_{t_i}^t \!\! dt_1 \!\! \int_{t_i}^{t_1} 
\!\!\! dt_2 \dots \!\int_{t_i}^{t_{n-1}} \!\!\!\!\!\!\!\!\! dt_n 
{\cal A}(t_1,k) \cdots {\cal A}(t_n,k) . \quad
\end{eqnarray}
The exponent matrix ${\cal A}(t,k)$ vanishes whenever $q(t)$ is constant. It 
has the form,
\begin{equation}
{\cal A}(t,k) \! = \! \!\frac{\pi}4 \dot{\nu} \left(\matrix{ \!\! {\rm 
csc}(\nu \pi) c_{\nu}({\scriptstyle \!-\! \frac{x}{q}}) \!\! & \!\! - 2 i 
d_{\nu}({\scriptstyle \!-\! \frac{x}{q}}) \!\! \cr \!\! -2 i {\rm csc}^2\!(\nu 
\pi) b_{\nu}({\scriptstyle \!-\! \frac{x}{q}}) \!\! & \!\! -{\rm csc}(\nu \pi) 
c_{\nu}({\scriptstyle \!-\! \frac{x}{q}}) \!\!} \right) ,
\end{equation}
where the various coefficient functions are,
\begin{eqnarray}
b_{\nu}(z) \!\!\! & = & \!\!\! {1 \over 2 \sqrt{\pi}} \sum_{n=1}^{\infty} 
{(-1)^n \Gamma(n \! - \! \nu \! - \! \frac12) z^{2n \! - \! 2\nu} (n \! - \!
\nu)^{- \! 1} \over \Gamma(n) \Gamma(n \!-\! \nu \!+\! 1) \Gamma(n \!-\! 2\nu 
\!+\! 1)} , \quad \label{bnu} \\
c_{\nu}(z) \!\!\! & = & \!\!\! - {4 \over \pi} \sin(\nu \pi) \Bigl[ \psi(\nu)
- 1 - \ln({\scriptstyle \frac12} z) \Bigr] \nonumber \\
& & \quad - {1 \over \sqrt{\pi}} \sum_{n=1}^{\infty} {(-1)^n \Gamma(n \!-\! 
\frac12) z^{2n} n^{-1} \over \Gamma(n \! + \! \nu) \Gamma(n \! + \! 1) 
\Gamma(n \! - \! \nu \! + \! 1) } , \quad \label{cnu} \\
d_{\nu}(z) \!\!\! & = & \!\!\! {1 \over 2 \sqrt{\pi}} \sum_{n=0}^{\infty} 
{(-1)^n \Gamma(n \! + \! \nu \! - \! \frac12) z^{2n \! + \! 2\nu} (n \! + \!
\nu)^{- \! 1} \over \Gamma(n \!+\! 2 \nu) \Gamma(n \!+\! \nu \!+\! 1) 
\Gamma(n \!+\! 1)} , 
\quad \label{dnu}
\end{eqnarray}
Here $\psi(z) \equiv \Gamma'(z)/\Gamma(z)$.

Equation (\ref{MMCSeqn}) is of phenomenological interest because it gives 
the mode functions of linearized gravitons \cite{Grish1}. However, the 
dominant perturbations imprinted in the cosmic microwave background seem 
to be from the scalar modes of gravity plus a minimally coupled scalar 
\cite{WMAP}. The perturbative background and the equations defining the
mode functions of this system are reviewed in Section 2. In Section 3 we 
solve for the mode functions by altering the scale factor and time parameter
in (\ref{genu}). In Section 4 the general solutions are re-expressed in 
terms of the actual scale factor and the co-moving time $t$. We give simple 
and accurate approximations for the ultraviolet regime ($x(t,k) \gg 1$) and 
the infrared regime ($x(t,k) \ll 1$) in Section 5. Section 6 discusses 
applications. 

{\it 2. Gravity with a minimally coupled scalar:} The Lagrangian in which
we are interested is,
\begin{equation}
{\cal L} = \frac1{16 \pi G} R \sqrt{-g} - \frac12 \partial_{\mu} \varphi 
\partial_{\nu} \varphi g^{\mu\nu} \sqrt{-g} - V(\varphi) \sqrt{-g} \; .
\end{equation}
This system can be solved perturbatively in a variety of different field 
variables and gauge conditions \cite{MFB}. We will use a recent formulation
\cite{ITTW,AW} based upon generalizing the de Donder gauge condition 
typically employed for quantum gravity in flat space. 

The full scalar and metric fields are expressed in terms of their 
background values plus quantum fields,
\begin{eqnarray}
\varphi(\eta,\vec{x}) & = & \varphi_0(\eta) + \phi(\eta,\vec{x}) \; , \\
g_{\mu\nu}(\eta,\vec{x}) & = & a^2 \Bigl( \eta_{\mu\nu} + \kappa 
\psi_{\mu\nu}(\eta,\vec{x}) \Bigr) \; .
\end{eqnarray}
Here $\kappa^2 \equiv 16 \pi G$ is the loop counting parameter of quantum 
gravity. As usual, indices on the graviton field $\psi_{\mu\nu}$ are raised 
and lowered using the Minkowski metric $\eta_{\mu\nu}$.

Note that we consider the fields to be functions of conformal time $\eta$, 
\begin{equation}
\eta \equiv \eta_i + \int_{t_i}^t \frac{dt'}{a(t')} \quad \Longrightarrow 
\quad \frac{d}{d\eta} = a(t) \frac{d}{dt} \; . \label{conf}
\end{equation}
This in no way precludes expressing the relations between the background
fields $\varphi_0(\eta)$ and $a(t)$ in terms of co-coving time derivatives,
\begin{eqnarray}
3 H^2 & = & 8 \pi G \Bigl( \frac12 \dot{\varphi}_0^2 + V(\varphi_0) \Bigr)
\; , \\
-2 \dot{H} - 3 H^2 & = & 8 \pi G \Bigl( \frac12 \dot{\varphi}_0^2 - 
V(\varphi_0) \Bigr) \; . 
\end{eqnarray}
As usual, an overdot denotes differentiation with respect to $t$ whereas
a prime stands for differentiation with respect to $\eta$.

It is traditional with this system to consider the scalar potential 
$V(\varphi)$ to be a known function from which the scale factor is inferred.
However, for our purposes it is more convenient to regard $a(t)$ as the 
known function. If one desires the scalar and its potential they can be 
reconstructed using the relations,
\begin{equation}
\dot{\varphi}_0 = - \frac{\sqrt{1\!+\!q}\, H}{\sqrt{4 \pi G}} \quad {\rm and} 
\quad V(\varphi_0) = \frac{(2\!-\!q) H^2}{8 \pi G} \; .
\end{equation}

The generalized de Donder gauge condition is \cite{ITTW},
\begin{equation}
F_{\mu} \equiv a \Bigl[ \psi^{\nu}_{~\mu , \nu} \!-\! \frac12 \psi^{\nu}_{~\nu 
, \mu} \!-\! 2 a H \psi_{\mu 0} \!+\! 2 \delta^0_{\mu} \sqrt{1\!+\! q} \, 
H a \phi \Bigl] = 0 \; . \label{dD}
\end{equation}
In this gauge it can be shown that the linearized fields of this system 
--- and the Green's functions needed to solve at higher orders --- are 
described by three plane wave mode functions $Q_I(\eta,k)$ \cite{AW},
\begin{equation}
Q_I^{\prime\prime} + \Bigl(k^2 - \frac{\theta_I^{\prime\prime}}{\theta_I}
\Bigr) Q_I = 0 \qquad {\rm for} \;\; I = A, B, C \; , \label{Qeqn}
\end{equation}
The three parameter functions $\theta_I$ are,
\begin{equation}
\theta_A = a \quad , \quad \theta_B = \frac1{a} \quad {\rm and} \quad 
\theta_C = \frac1{\sqrt{1\!+\!q}\, a} \; . \label{thetas}
\end{equation}
We normalize the various mode functions to obey,
\begin{equation}
Q_I Q_I^{*\prime} - Q_I^{\prime} Q_I^* = i \; .
\end{equation}

The linearized fields in de Donder gauge (\ref{dD}) can be expressed
using two mode sums \cite{AW},
\begin{eqnarray}
\psi_{00} & \!\!=\!\! & -\sqrt{1\!+\!q} \, H \!\!\int\!\! \frac{d^3k}{(2\pi)^3} 
\Bigl\{ \frac1{k} Q_C e^{i \vec{k} \cdot \vec{x}} Y(\vec{k}) + {\rm c.c.} 
\Bigr\} , \quad \\
\psi^{tt}_{ij} & \!\!=\!\! & \frac1{a} \!\! \int \!\! \frac{d^3k}{(2\pi)^3} \! 
\sum_{\lambda} \Bigl\{ \epsilon_{ij}(\vec{k},\lambda) Q_A e^{i \vec{k} 
\cdot \vec{x}} \alpha(\vec{k},\lambda) + {\rm c.c.} \Bigr\} . \qquad
\end{eqnarray}
Note that in these mode sums we have suppressed the arguments of previously 
defined functions such as $a(t)$, $H(t)$, $q(t)$ and $Q_I(\eta,k)$. The 
scalar and graviton creation and annihilation operators are canonically 
normalized,
\begin{eqnarray}
\Bigl[ Y(\vec{k}) \, , \, Y^{\dagger}(\vec{k}') \Bigr] & = & (2\pi)^3 
\delta^3(\vec{k} - \vec{k}') \; , \\
\Bigl[ \alpha(\vec{k},\lambda) \, , \, \alpha^{\dagger}(\vec{k}',\lambda') 
\Bigr] & = & (2\pi)^3 \delta^3(\vec{k} - \vec{k}') \, \delta_{\lambda 
\lambda'} \; .
\end{eqnarray}
As usual, graviton polarization tensors are transverse ($k_i \epsilon_{ij}(
\vec{k},\lambda) = 0$), traceless ($\delta_{ij} \epsilon_{ij}(\vec{k},\lambda) 
= 0$), and orthonormal ($\epsilon_{ij}(\vec{k},\lambda) \epsilon^*_{ij}(
\vec{k},\lambda') = \delta_{\lambda \lambda'}$). The other linearized fields 
are,
\begin{eqnarray}
\phi(\eta,\vec{x}) & = & \frac1{\sqrt{1\!+\!q} \, H a} \frac{\partial}{
\partial t} \Bigl( a \psi_{00}(\eta,\vec{x}) \Bigr) \; , \\
\psi_{oi}(\eta,\vec{x}) & = & 0 \; , \\
\psi_{ij}(\eta,\vec{x}) & = & \delta_{ij} \psi_{00}(\eta,\vec{x}) + 
\psi^{tt}_{ij}(\eta,\vec{x}) \; .
\end{eqnarray}
Although the $B$ modes are not required at linearized order it is no trouble
to get them, and they would be appear in the retarded Green's functions 
needed to obtain the next order solutions \cite{AW}.

{\it 3. Reduction to the MMC scalar mode functions:} Our procedure is to 
define a new scale factor and a new time coordinate so that the $Q_I$ mode
equation (\ref{Qeqn}) reduces to the mode equation (\ref{MMCSeqn}) of the
massless, minimally coupled scalar. There does not seem to be any 
way to motivate why this works, but the trick is simple enough that we 
shall simply state it. 

We define the new time variable $t_I$ as follows,
\begin{equation}
dt_I = \pm \theta_I d\eta \quad \Longrightarrow \quad \frac{d}{d\eta} = \pm 
\theta_I \frac{d}{dt_I} \; . \label{TI}
\end{equation}
The plus sign is used for functions $\theta_I$ which grow with conventional 
time (hence $I = A$) while the minus is taken when $\theta_I$ falls (hence 
$I = B,C$). Note that $\theta_I$ plays the same role with respect to $\pm 
t_I$ as the scale factor $a(t)$ does with respect to the co-moving time $t$. 
This suggests that we define associated Hubble and deceleration parameters,
\begin{equation}
H_I \equiv \frac1{\theta_I} \frac{d \theta_I}{d t_I} \quad {\rm and} \quad
q_I \equiv -1 -\frac1{H_I^2} \frac{d H_I}{d t_I} \; . \label{Igeom}
\end{equation}
A little calculus establishes the identity,
\begin{equation}
\Bigl( \frac{d}{d \eta} \Bigr)^2 = \theta_I^2 \Bigl[ \Bigl( \frac{d}{d t_I}
\Bigr)^2 + H_I \frac{d}{d t_I} \Bigr] \; .
\end{equation}
Applying this to the function $\theta_I$ itself we can re-express the 
quantity $-\theta_I^{\prime\prime}/\theta_I$ which appears in the $Q_I$ 
mode equation (\ref{Qeqn}),
\begin{equation}
-\frac{\theta_I^{\prime\prime}}{\theta_I} = \theta_I^2 H_I^2 ( -1 + q_I) \; .
\end{equation}
In keeping with the geometrical analogy, we define the ``physical'' wave 
number in ``Hubble'' units as,
\begin{equation}
x_I \equiv \frac{k}{\theta_I H_I} \; .
\end{equation}

The final step is to rescale $Q_I$ by a factor of $\sqrt{\theta_I}$,
\begin{equation}
u_I \equiv \sqrt{\theta_I} Q_I \; \Longrightarrow \; \frac{d^2 u_I}{d t_I^2}
+ H_I^2 \Bigl[ x_I^2 + \frac32 q_I - \frac34 \Bigr] u_I = 0 \; . \label{key}
\end{equation}
This is the same equation for a plane wave mode function $u_I$ 
evolving in the geometry $ds_I^2 = -dt_I^2 + \theta_I^2 d\vec{x} 
\cdot d\vec{x}$ as (\ref{MMCSeqn}) is for the mode function $u(t,k)$ 
evolving in the geometry $ds^2 = -dt^2 + a^2 d\vec{x} \cdot d\vec{x}$.
Even the Wronskian agrees up to a sign,
\begin{equation}
u_I \frac{d u_I^*}{d t_I} - \frac{d u_I}{d t_I} u_I^* = \pm \Bigl[
Q_I Q_I^{*\prime} - Q_I^{\prime} Q_I^* \Bigr] = \pm i \; .
\end{equation}
We can absorb the sign by making the $B$ and $C$ modes scale to 
$Q^*$, which does not change (\ref{key}),
\begin{eqnarray}
u_A(t_A,k) & \equiv & \sqrt{\theta_A} \, Q_A(\eta,k) \; , \label{defuA} \\
u_B(t_B,k) & \equiv & \sqrt{\theta_B} \, Q_B^*(\eta,k) \; , \label{defuB} \\
u_C(t_C,k) & \equiv & \sqrt{\theta_C} \, Q_C^*(\eta,k) \; . \label{defuC}
\end{eqnarray}
It follows that $u_I$ must be the same function of $t_I$, with scale 
factor $\theta_I$, as $u(t,k)$ is of $t$, with the scale factor $a(t)$.

{\it 4. Expressing the solution in conventional form:} Of course 
physics is not based on the fictitious geometry $ds_I^2 = -dt_I^2 + 
\theta_I^2 d\vec{x} \cdot d\vec{x}$, so we must express the solutions 
$u_I$ as functions of co-moving time $t$ and as functionals of the true 
scale factor $a(t)$. From their definitions (\ref{thetas}) we already 
know how the functions $\theta_I(t)$ depend upon the co-moving time. 
To determine $H_I(t)$ and $q_I(t)$ we need the differential relation 
between $t_I$ and $t$. By comparing the transformation (\ref{TI}) from 
$\eta$ to $t_I$ with the transformation (\ref{conf}) from $\eta$ to $t$ 
we infer,
\begin{equation}
\frac{d}{d\eta} = a \frac{d}{dt} \quad \Longrightarrow \quad \frac{d}{d t_I} 
= \pm \frac{a}{\theta_I} \frac{d}{dt} \; . \label{tItot}
\end{equation}
As before, the $+$ sign pertains for $A$ modes and the $-$ sign for 
$B$ and $C$ modes.

Now apply (\ref{tItot}) to the parameters (\ref{Igeom}) of the $ds_I^2$
geometry. The $A$ and $B$ mode parameters are simple,
\begin{eqnarray}
H_A = H & , & q_A = q \; , \\
H_B = H a^2 & , & q_B = - q \; .
\end{eqnarray}
The $C$ mode parameters are more complicated,
\begin{equation}
H_C = (1\!+\!r) \sqrt{1\!+\!q}\, H a^2 \; , \; q_C = - \frac{q}{1\!+\!r}
+ \frac{\frac{\dot{r}}{H}}{(1\!+\!r)^2} \; ,
\end{equation}
where we define the additional parameter,
\begin{equation}
r(t) \equiv \frac1{H} \frac{d}{dt} \ln\Bigl(\sqrt{1\!+\!q}\Bigr) \; .
\end{equation}
However, note that whenever $q(t)$ is constant --- which is the case for all
familiar cosmologies --- the $C$ mode parameters degenerate,
\begin{equation}
\dot{q} = 0 \; \Longrightarrow \; H_C = \sqrt{1\!+\!q}\, H a^2 \;\; {\rm and} 
\;\; q_C = - q \; .
\end{equation}
The rescaled wave number is in all cases simple,
\begin{equation}
x_A = x(t,k) \; , \; x_B = x(t,k) \;\; {\rm and} \;\; x_C = \frac{x(t,k)}{1
\!+\! r(t)} \; .
\end{equation}
In fact, whenever $q(t)$ is constant, $x_C = x(t,k)$.

We have reached the point where the parameters of the $ds_I^2$ geometry
can all be considered to be functions of co-moving time $t$: $\theta_I(t)$, 
$H_I(t)$, $q_I(t)$ and $x_I(t,k)$. We make the additional definition 
$\nu_I(t) \equiv \frac12 \!-\! q^{\!-\!1}_I(t)$. Based on the work of the 
previous section the $I=A,B,C$ mode functions can be read off from 
(\ref{genu}) by replacing the parameters of the $ds^2$ geometry with the
corresponding $ds_I^2$ parameters, considered as functions of $t$,
\begin{eqnarray}
\lefteqn{u_I(t,k) \! \equiv \! \sqrt{\frac{\pi \theta_I}{2k}} \Bigl(\!-\!
\frac{x_I}{2q_I} \Bigr)^{\!\frac12} \!\Bigl(\!-i {\rm csc}(\nu_I \pi) J_{\!-\!
\nu_I}\!({\scriptstyle \!-\! \frac{x_I}{q_I}}) \, , \, J_{\nu_I}\!({
\scriptstyle \!-\! \frac{x_I}{q_I}}) \Bigr) } \nonumber \\
& & \hspace{2cm} \times {\cal M}_I(t,t_i,k) \left(\matrix{1 \cr 1 \!+\! i 
{\rm cot}[\nu_I(t_i) \pi] }\right) . \qquad \label{genuI}
\end{eqnarray}
As with $u(t,k)$, the transfer matrix takes the form,
\begin{equation}
{\cal M}_I(t,t_i,k) \equiv P\left\{ \exp\left[ \int_{t_i}^t dt' 
{\cal A}_I(t',k) \right] \right\} \; .
\end{equation}
Its exponent matrix obeys the same scheme,
\begin{equation}
\! {\cal A}_I\!(t,k) \! = \! \!\frac{\pi}4 \dot{\nu}_I \!\!\left(\matrix{ 
\!\! {\rm csc}(\nu_I \pi) c_{\nu_I}\!(\!{\scriptstyle \!-\! \frac{x_I}{q_I}}\!)
\!\! & \!\! - 2 i d_{\nu_I}\!({\scriptstyle \!-\! \frac{x_I}{q_I}}\!) \!\! \cr 
\!\! -2 i {\rm csc}^2\!(\nu_I \pi) b_{\nu_I}\!(\!{\scriptstyle \!-\!
\frac{x_I}{q_I}}\!) \!\! & \!\! -{\rm csc}(\nu_I \pi) c_{\nu_I}\!(\!{
\scriptstyle \!-\! \frac{x_I}{q_I}}\!) \!\!} \right) \!\! ,
\end{equation}
where the coefficient functions (\ref{bnu}-\ref{dnu}) are unchanged. Note
that we were able to express the transfer matrix in terms of integrations
over the co-moving time owing to the relation,
\begin{equation}
\int_{t_I(t_1)}^{t_I(t_2)} \!\!\!\!\!\!\!\!\! dt_I \frac{d \nu_I}{d t_I} 
F(x_I,q_I) = \int_{t_1}^{t_2} \!\!\!\!\!\! dt \, \dot{\nu_I} F(x_I,q_I) \; .
\end{equation}

{\it 5. The ultraviolet and infrared regimes:} The ultraviolet regime is
defined by $x(t,k) \gg 1$. In this limit we know from previous work 
\cite{TW1} that the three mode functions approach the form,
\begin{equation}
u_I(t,k) \Bigl\vert_{x\gg1} \!\!\! \longrightarrow \sqrt{\frac{\theta_I}{2 k}}
\exp\Bigl[ i \chi_I(t_i,k) \!-\! i k \!\! \int_{t_I(t_i)}^{t_I(t)} \!
\frac{d t_I'}{\theta_I} \Bigr] ,
\end{equation}
where the phase $\chi_I$ is,
\begin{equation}
\chi_I(t,k) \equiv - \frac{x_I(t,k)}{q_I(t)} + \frac{1 \!-\! q_I(t)}{q_I(t)}
\, \frac{\pi}{2} \; .
\end{equation}
Recalling $dt_I = \pm \theta_I dt/a$, and the definition (\ref{conf}) of 
conformal time, we see that the various $Q_I$'s approach,
\begin{equation}
Q_I(\eta,k) \Bigl\vert_{x \gg 1} \longrightarrow \frac1{\sqrt{2k}}
e^{-i k (\eta \!-\! \eta_i) \pm i \chi_I(t_i,k)} \; .
\end{equation}
Note that the $\pm$ signs in the relation between $dt_I$ and $dt$ 
exactly cancels the variations (\ref{defuA}-\ref{defuC}) in the relations 
between $u_I(t,k)$ and $Q_I(\eta,k)$.

The infrared regime ($x \ll 1$) is more subtle and more interesting. We 
assume that the wave number was initially ultraviolet, i.e., $x(t_i,k)
\gg 1$. During inflation $x(t,k)$ falls, typically exponentially. This
drives the lower ultraviolet modes through first horizon crossing ($x(t_1,k)
= 1$) into the infrared regime of $x(t,k) \ll 1$. In this regime one of the
Bessel functions will enormously dominate the other. It is therefore good
to re-express the general solution in terms of solutions which are the pure
$\pm \nu_I$ Bessel functions at first horizon crossing,
\begin{eqnarray}
\lefteqn{u_I(t,k) \equiv  \frac1{\sqrt{2}} \Bigl( u^-_I(t,t_1,k) \, ,
\, u^+_I(t,t_1,k)  \Bigr) } \nonumber \\
& & \hspace{2cm} \times {\cal M}_I(t_1,t_i,k) \left(\matrix{1 \cr 1 \!+\! i 
{\rm cot}[\nu_I(t_i) \pi] }\right) . \qquad 
\end{eqnarray}
Here $u^{\pm}_I(t,t_1,k)$ are the pure $J_{\pm\nu_I}$ solutions at first 
horizon crossing, evolved to time $t > t_1$.

As long as the condition $x_I(t,k) \ll 1$ is satified, the transfer matrix
can be worked out up to negligible correction factors of order $x^2$. To
the same fractional error we can also use the leading terms in the series 
expansions of the Bessel functions. The resulting form for the $-$ solution
is,
\begin{equation}
u^-_I(t,k) \Bigl\vert_{x \ll 1} \longrightarrow \frac{-i 
\theta_I^{\frac32}\!(t)}{\sqrt{\pi k}} \left[\frac{\Gamma(\frac12 \!-\!
\frac1{q_I})}{\theta_I \, (\frac{-x_I}{2 q_I})^{\!-\!\frac{1}{q_I}}}
\right]_{t_1} .
\end{equation}
The $+$ solution involves an integral,
\begin{eqnarray}
\lefteqn{u^+_I(t,k) \Bigl\vert_{x \ll 1} \longrightarrow \sqrt{\pi k} \,
\theta_I^{\frac32}\!(t) \left[\frac{\theta_I \, (\frac{-x_I}{2 q_I})^{\!-\!
\frac{1}{q_I}}}{\Gamma(\frac12 \!-\! \frac1{q_I})} \right]_{t_1} } \nonumber \\
& & \hspace{1cm} \times \left\{ \left[\frac{-1}{2 \nu_I q_I H_I \theta_I^3}
\right]_{t_1} \mp \int_{t_1}^t \!\! \frac{dt'}{a(t') \theta_I^2(t')}
\right\} .
\end{eqnarray}

For the $A$ mode it is the $u_A^-$ solution that dominates. We express 
$u_A(t,k)$ with the standard normalization times two $k$-dependent correction 
factors,
\begin{equation}
u_A(t,k) \Bigl\vert_{x \ll 1} \!\!\!\longrightarrow \frac{-i H_1}{2 k^3}
\, a^{\frac32}\!(t) \, {\cal C}_{1A}(k) \, {\cal C}_{iA}(k) \; . 
\label{IRuA}
\end{equation}
Here $H_1 \equiv H(t_1)$ is the Hubble parameter at first horizon crossing.
Note that it can depend upon $k$ because the time of first horizon crossing
depends upon the wave number. Indeed, some modes never experience horizon
crossing.

The correction factor in (\ref{IRuA}) that depends upon the system's state 
at first horizon crossing is,
\begin{equation}
{\cal C}_{1A}(k) \equiv \frac{\frac1{\sqrt{\pi}} \Gamma(\frac12 \!-\!
\frac1{q_1})}{(\!-\! \frac1{2 q_1})^{\!-\!\frac{1}{q_1}}} \; .
\end{equation}
Here $q_1 \equiv q(t_1)$ is the deceleration parameter at first horizon 
crossing. The correction factor depending upon evolution from $t_i$ to 
$t_1$ is,
\begin{equation}
{\cal C}_{iA}(k) \equiv {\cal M}^{11}_A(t_1,t_i,k) + {\cal M}^{12}_A(t_1,t_i,k)
\, e^{i \frac{\pi}{q_i}} {\rm sec}({\scriptstyle \frac{\pi}{q_i}}) \; .
\end{equation}
Here $q_i \equiv q(t_i)$ is the initial value of the deceleration parameter.
Since $\dot{\nu}(t) = \frac{\dot{q}(t)}{q^2(t)}$ is typically small during
inflation, it ought to be a very good approximation to simply take the first
several terms of the series expansion of the transfer matrix in estimating 
${\cal C}_{iA}(k)$. Combining (\ref{IRuA}) and (\ref{defuA}) we see that the 
infrared limit of $Q_A$ takes the form,
\begin{equation}
Q_A(\eta,k) \Bigl\vert_{x \ll 1} \!\!\! \longrightarrow \frac{-i H_1}{ 
\sqrt{2 k^3}} \, a(t) \, {\cal C}_{1A}(k) \, {\cal C}_{iA}(k) \; .
\end{equation}

For the $B$ mode it is the $u_B^+$ solution that dominates, and only the 
integral is significant,
\begin{equation}
\! u_B(t,k) \Bigl\vert_{x \ll 1} \!\!\! \longrightarrow 
\frac{\; H_1}{\sqrt{2 k}} \frac1{a^{\!\frac32}\!(t)} \! \int_{t_1}^t \!\! dt' 
a(t') \, {\cal C}_{1B}(k) \, {\cal C}_{i B}(k) . \label{IRuB}
\end{equation}
The correction factor from horizon crossing is,
\begin{equation}
{\cal C}_{1 B}(k) \equiv \frac{\frac1{\sqrt{\pi}} \Gamma(\frac12 \!-\!
\frac1{q_1})}{(\!-\! \frac1{2 q_1})^{\!-\!\frac{1}{q_1}}} \, 
e^{i \frac{\pi}{q_1}} \cos({\scriptstyle \frac{\pi}{q_1}}) \; .
\end{equation}
The other correction factor depends upon evolution up to this time,
\begin{equation}
{\cal C}_{i B}(k) \equiv {\cal M}^{21}_B(t_1,t_i,k) + {\cal M}^{22}_B(t_1,
t_i,k) \, e^{-i \frac{\pi}{q_i}} {\rm sec}({\scriptstyle \frac{\pi}{q_i}}) \; .
\end{equation}
And we can combine (\ref{IRuB}) and (\ref{defuB}) to extract the infrared 
limit of $Q_B$,
\begin{equation}
\! Q_B(\eta,k) \Bigl\vert_{x \ll 1} \!\!\! \longrightarrow 
\frac{\; H_1}{\sqrt{2k}} \frac1{a(t)} \! \int_{t_1}^t \!\! dt' a(t') \, 
{\cal C}^*_{1B}(k) \, {\cal C}^*_{i B}(k) .
\end{equation}

It is the $u_C^+$ solution that dominates the infrared $C$ mode,
\begin{eqnarray}
\lefteqn{u_C(t,k) \Bigl\vert_{x \ll 1} \!\!\! \longrightarrow 
\frac{\; H_1}{\sqrt{2 k (1 \!+\! q_1)}} \frac1{[1\!+\!q(t)]^{\frac34}} 
\frac1{a^{\!\frac32}\!(t)} } \nonumber \\
& & \hspace{1.5cm} \times \int_{t_1}^t \!\! dt' a(t') [1 + q(t')] \, 
{\cal C}_{1 C}(k) \, {\cal C}_{i C}(k) . \qquad \label{IRuC}
\end{eqnarray}
The correction factor at horizon crossing takes the form,
\begin{equation}
{\cal C}_{1 C}(k) \equiv \frac{\frac1{\sqrt{\pi}} \Gamma(\frac12 \!+\!
\frac1{q_{1C}})}{(\frac1{2 q_{1C}})^{\frac{1}{q_{1C}}}} \, e^{-i 
\frac{\pi}{q_{1C}}} \cos({\scriptstyle \frac{\pi}{q_{1C}}}) (1 + r_1)^{\frac1{
q_{1C}}} . \label{C1C}
\end{equation}
where $q_{1C} \equiv q_C(t_1)$ and $r_1 \equiv r(t_1)$. We remind the reader 
that the $C$ deceleration parameter and $r(t)$ are,
\begin{eqnarray}
q_C(t) & = & - \frac{q(t)}{1 \!+\! r(t)} + \frac{\frac{\dot{r}(t)}{H(t)}}{[1 
+ r(t)]^2} \; , \label{qC} \\
r(t) & = & \frac1{H(t)} \frac{d}{dt} \ln\Bigl(\sqrt{1 + q(t)} \,\Bigr) \; .
\end{eqnarray}
The other correction factor depends upon $q_{iC} \equiv q_C(t_i)$,
\begin{equation}
{\cal C}_{i C}(k) \equiv {\cal M}^{21}_C(t_1,t_i,k) + {\cal M}^{22}_C(t_1,
t_i,k) \, e^{-i \frac{\pi}{q_{iC}}} {\rm sec}({\scriptstyle 
\frac{\pi}{q_{iC}}}) \; . \label{CiC}
\end{equation}
Taking (\ref{IRuC}) together with (\ref{defuC}) gives the last of our
infrared limits,
\begin{eqnarray}
\lefteqn{Q_C(\eta,k) \Bigl\vert_{x \ll 1} \!\!\! \longrightarrow 
\frac{\; H_1}{\sqrt{2k (1 \!+\! q_1)}} \frac1{\sqrt{1\!+\!q(t)}} \frac1{a(t)} }
\nonumber \\
& & \hspace{1.5cm} \times \int_{t_1}^t \!\! dt' a(t') [1 + q(t')] \, 
{\cal C}^*_{1 C}(k) \, {\cal C}^*_{i C}(k) . \qquad \label{IRQC}
\end{eqnarray}

{\it 6. Applications:} The formalism we have developed might seem intimidating, 
and one must of course approximate the transfer matrix in order to compute 
anything. However, it is often of great value to possess general 
expressions. For example, one can search for unexpected effects such as the 
consequences of the universe having passed through $q = 0$ at the end of
inflation. We can also explore cosmologies for which the $C$ mode deceleration 
parameter (\ref{qC}) differs appreciably from its small $r(t)$ limit of
$-q(t)$.

Our formalism is also useful. For example, our infrared limiting form 
(\ref{IRQC}) allows one to compute the Sachs-Wolfe contribution to the 
primordial scalar power spectrum for any evolution $a(t)$ which does not 
compromise the infrared limit. It should be noted, for non-experts, that
this primordial power spectrum is almost perfectly flat. The rich structure
imaged by WMAP \cite{WMAP} derives from a variety of late-time effects 
which occur after second horizon crossing, when the stress-energy is no
longer dominated by the inflaton and our scalar mode functions are not
relevant. (However, note that our tensor mode functions continue to apply.
They depend only upon $a(t)$, not what caused it.) The primordial power
spectrum sets one of the initial conditions for this late-time cosmology,
whose dynamics are well understood.

Assuming the cosmic microwave radiation radiation was emitted during pure 
matter domination ($q = +\frac12$), our result for the primordial scalar
power spectrum is \cite{TW2},
\begin{equation}
{\cal P}_{\rm SW}(k) = \frac{9}{4\pi} \frac{G H_1^2}{1\!+\! q_1} \Vert
{\cal C}_{1C}(k) \Vert^2 \Vert {\cal C}_{iC}(k) \Vert^2 \; . \label{PSW}
\end{equation}
The standard result is $\frac9{4\pi} G H_1^2/(1\!+\!q_1)$ so the correction
factors (\ref{C1C}) and (\ref{CiC}) represent improvements. Because
different conventions exist in the literature we correspond ${\cal P}_{\rm 
SW}(k)$ below to the symbol $\delta(k)$ used by Mukhanov, Feldman and 
Brandenberger \cite{MFB}, to the variable ${\cal P}_{\cal R}(k)$ used by 
Liddle and Lyth \cite{LL}, and to the quantity $A_S^2(k)$ used by Lidsey 
{\it et al.} \cite{Rocky},
\begin{equation}
{\cal P}_{\rm SW}(k) = \frac{25}4 \Vert \delta(k) \Vert^2 = \frac94 
{\cal P}_{\cal R}(k) = \frac{225}{16} A_S^2(k) \; .
\end{equation}
We also specify how it enters the correlation function between temperature 
fluctuations observed from directions $\widehat{e}_1$ and $\widehat{e}_2$,
\begin{eqnarray}
\lefteqn{\Bigl\langle \Omega \Bigl\vert \frac{\Delta T_R(\widehat{e}_1)}{
T_R} \frac{\Delta T_R(\widehat{e}_2)}{T_R} \Bigr\vert \Omega 
\Bigr\rangle_{\rm SW} } \nonumber \\
& & \hspace{1cm} = \!\! \int_0^{\infty} \!\! \frac{dk}k {\cal P}_{\rm SW}(k) 
\int \!\! \frac{d^2 \widehat{k}}{4 \pi} e^{2 i x_0 \widehat{k} \cdot 
(\widehat{e}_1 \!-\! \widehat{e}_2)} ,
\end{eqnarray}
where $x_0 \equiv x(t_0,k)$ is the physical wave number in current Hubble
units.

A very interesting problem we now have the analytical power to tackle is
the apparent paradox associated with the fact that the scalar power spectrum
(\ref{PSW}) diverges whenever first horizon crossing occurs during a 
temporary period of de Sitter inflation (i.e., $q_1 = -1$). It is difficult
to understand why the result in this case is not of gravitational strength, 
and one would think it should not diverge. This has led to controversy in the
past \cite{Grish2} which it is now possible to resolve.

Finally, these mode functions can be used to study back-reaction in the 
gravity plus scalar system. It has been established that no secular 
back-reaction occurs at one loop for simple models \cite{AW,GB}. This 
formalism allows one to study an arbitrary single-scalar model of inflation.

{\it 7. Acknowledgements:} This work was partially supported by EU
grants HPRN-CT-2000-00122 and HPRN-CT-2000-00131, by DOE contract 
DE-FG02-97ER41029 and by the Institute for Fundamental Theory.


\end{document}